\documentclass[twocolumn,prb,printnumbers,superscriptaddress,amsmath,amssymb]{revtex4}
\usepackage{graphicx}
\usepackage{color}

\begin{document}

\title{Sound attenuation in stable glasses}
\date{today}

\author{Lijin Wang}
\affiliation{Beijing Computational Science Research Center, Beijing 100193, P. R. China}
\affiliation{Department of Chemistry, Colorado State University, Fort Collins, Colorado 80523, USA}
\author{Ludovic Berthier}
\affiliation{Laboratoire Charles Coulomb (L2C), University of Montpellier, CNRS, 34095 Montpellier, France}
\author{Elijah Flenner}
\affiliation{Department of Chemistry, Colorado State University, Fort Collins, Colorado 80523, USA}
\author{Pengfei Guan}
\affiliation{Beijing Computational Science Research Center, Beijing 100193, P. R. China}
\author{Grzegorz Szamel}
\affiliation{Department of Chemistry, Colorado State University, Fort Collins, Colorado 80523, USA}

\begin{abstract}
Understanding the difference between universal low-temperature properties of amorphous and crystalline solids requires an explanation
of the stronger damping of long-wavelength phonons in amorphous solids.
A longstanding sound attenuation scenario, resulting from a combination of experiments,
theories, and simulations, leads to a quartic scaling of sound attenuation with the wavevector,
which is commonly attributed to Rayleigh scattering of the sound.
Modern computer simulations offer conflicting conclusions regarding the validity of this picture.
We simulate glasses with an unprecedentedly broad range of stabilities to perform the first microscopic
analysis of sound damping in model glass formers across a range of experimentally relevant preparation protocols.
We present a  convincing evidence  that quartic scaling is recovered for small wavevectors irrespective of the glass's stability.
With increasing stability, the wavevector where the quartic scaling begins increases by approximately a factor of three
and the sound attenuation decreases
by over an order of magnitude. Our results uncover an intimate connection between glass stability and sound damping.

\end{abstract}

\maketitle

\section{Introduction}

Many theoretical descriptions of sound attenuation in low temperature (athermal) amorphous solids predict
a quartic scaling of the sound attenuation with the wavevector. Early arguments, used to
explain the plateau in the temperature dependence of the thermal conductivity \cite{Zeller1971,Zaitlin1975},
invoked the picture of scattering of sound waves by uncorrelated inhomogeneities that are much smaller than the wavelength,
which is the physical scenario known as the Rayleigh scattering.  In several theories, these inhomogeneities have been modeled
as local fluctuations of elastic
constants~\cite{Schirmacher2015,Schirmacher2007,Schirmacher2010,MarruzzoScientific2013,Schirmacher2008,Schirmacher2006}. These
theories predict that the sound attenuation scales with the fourth power of the wavevector, $\Gamma_\lambda(k) \sim k^4$
($\lambda = L$ denotes longitudinal waves and $\lambda = T$ denotes transverse waves) for small wavevector $k$.
Mean-field theories~\cite{Wyart2010,DeGiuli2014,Vitelli2010,Grigera2011} arrive at the same prediction, albeit in a different way.
Yet another theoretical treatment, the soft-potential model, predicts that a quartic scaling regime exists due to phonons interacting
with soft modes~\cite{Buchenau1992}.

Longitudinal sound attenuation can be directly obtained from X-ray and light scattering experiments. A compilation of many experimental
results~\cite{Mizuno2014PNAS,RuoccoPRL1999,SetteScience1998,RuoccoJPCM2001,MasciovecchioPRL2004,ScopignoPRL2006,BenassiPRB2005,
MasciovecchioPRL2006,Rayleigh_RufflePRL2006,Rayleigh_MonacoPNAS2009_exp,Rayleigh_BaldiPRL2010,Rayleigh_BaldiPRL2014,
Rayleigh_BaldiPRL2013,Rayleigh_RutaJCP2012,Devos2008} shows that the wavevector dependence of the longitudinal sound attenuation
parameter, $\Gamma_{L}(k)$, can be divided into three regimes: (1) $\Gamma_L(k) \sim k^2$ for low $k$; (2)  $\Gamma_L(k) \sim k^4$
for an intermediate $k$ regime; and (3) $\Gamma_L(k) \sim k^2$ for large $k$. While the intermediate wavevector quartic and the
large wavevector quadratic scalings of the sound attenuation parameter are well-documented, the small wavevector quadratic
dependence was only seen in a few experiments~\cite{MasciovecchioPRL2004,MasciovecchioPRL2006,BenassiPRB2005,Devos2008}. Because
the experiments are performed at finite temperature, the small wavevector quadratic scaling can be ascribed to thermal and anharmonic
effects.

Computer simulations offer a conflicting view of these results. Most computer studies investigate sound attenuation
in the limit of zero temperature, in order to remove anharmonic effects. To our knowledge, no simulation reproduced
the $\Gamma_L(k) \sim k^2$ scaling observed at small wavevectors in
experiments~\cite{MasciovecchioPRL2004,MasciovecchioPRL2006,BenassiPRB2005,Devos2008}. Regarding the quartic Rayleigh scattering
regime, no firm conclusion can be drawn either. By simulating large glasses created by quenching configurations from a mildly
supercooled liquid, Gelin {\it et al.}~\cite{GelinNatMat2016} found a logarithmic correction to the quartic scaling,
$\Gamma_\lambda(k) \sim k^4 \ln(k)$. They invoked the existence of correlated inhomogeneities of the elastic
constants~\cite{John1983} to rationalize this observation. However, a more recent, larger-scale
study~\cite{MizunoArxiv2018phonontransport} of harmonic spheres close to their unjamming transition confirmed the Rayleigh scattering
scenario in 2D glasses and conjectured its validity in 3D glasses.
Finally, a very recent preprint~\cite{Lerner2019} (which appeared when the present paper was being finalized for submission)
presented the first convincing evidence of the small wavevector quartic scaling of the
transverse sound attenuation in a 3D glass created by quenching from a mildly supercooled liquid.
However, the status of the longitudinal sound attenuation, even for simple
glass-formers in the zero-temperature harmonic limit, remains unsettled.

To our knowledge, all prior simulations investigated sound attenuation in glasses with stabilities dramatically different from the ones of typical laboratory glasses, preventing direct comparison between numerics and real materials. This constraint is imposed by the large preparation times
required to equilibrate systems close to the experimental glass transition, which, therefore, cannot be simulated using conventional
techniques. In this work, we use an efficient swap Monte-Carlo algorithm~\cite{swap2001} that was recently
developed~\cite{swap2016,swapPRX2017} to prepare glasses with stabilities comparable to, or even exceeding, the
stability of experimental glasses. If we quantify the glass stability in terms of a cooling rate, the improvement due to the
swap algorithm is equivalent to decreasing the cooling rate by more than 10 orders of magnitude, thus closing the gap between previous
computer investigations and realistic materials. In previous studies, it was demonstrated that both the low-frequency vibrational
properties~\cite{WangarXiv2018} and the mechanical properties~\cite{yielding} of computer generated glasses dramatically evolve with
increasing the glass stability over such a broad range.

We find that changing the glass stability over a broad range fully clarifies the elusive picture of sound attenuation.
Generally, sound attenuation decreases with increasing stability, implying that more stable glasses are also less dissipative solids
(classical zero temperature crystalline solids are non-dissipative). More importantly, we find the wavevector dependence of sound
attenuation at low wavevectors exhibits a quartic scaling,  for both transverse and longitudinal modes and
in glasses with very different stabilities. Thus, we unambiguously demonstrate the universality of the Rayleigh scattering scaling
in 3D glasses. The quartic scaling of the sound attenuation with the wavevector is more prominent in more stable glasses,
which adds to the conjectured connection between glass stability and sound damping.

\section{Methods}
\subsection{Simulation details}

We perform computer simulations using a three-dimensional cubic system composed of polydisperse particles
with equal mass $m=1$. The distribution of particle diameters $\sigma$  follows
$P(\sigma)=\frac{A}{\sigma^3}$, where $\sigma\in [0.73,1.63]$ and $A$ is a normalization factor.
The cross-diameter $\sigma_{ij}$ is determined according to a non-additive mixing rule,
$\sigma_{ij}=\frac{\sigma_{i}+\sigma_{j}}{2}(1-\epsilon |\sigma_{i}-\sigma_{j}|)$ with $\epsilon=0.2$.
 The interaction between two particles ${i}$ and ${j}$  is given by
the inverse power law potential,
$ V(r_{ij}) = \left(\frac{\sigma_{ij}}{r_{ij}}\right)^{12} + V_{cut}(r_{ij})$,
when the separation $r_{ij}$ is smaller than the potential cutoff $r_{ij}^c=1.25\sigma_{ij}$,
and zero otherwise. Here,  $V_{cut}(r_{ij})=c_{0}+c_{2}\left(\frac{r_{ij}}{\sigma_{ij}}\right)^{2}
+ c_{4}\left(\frac{r_{ij}}{\sigma_{ij}}\right)^{4}$, and  the  coefficients $c_{0}$, $c_{2}$
and $c_{4}$ are set to guarantee the continuity of $V(r_{ij})$ at $r_{ij}^c$ up to the second
derivative.

We produce zero-temperature glasses by instantaneously quenching   supercooled liquids
equilibrated through the swap Monte Carlo algorithm  at different parent temperatures $T_p$, which uniquely control their stability~\cite{WangarXiv2018}, to their
local potential minima using the fast inertia relaxation engine minimization \cite{method_only_fire}.
We calculate the normal modes by diagonalizing the dynamic  matrix using Intel Math Kernel Library
(https://software.intel.com/en-us/mkl/) and ARPACK (http://www.caam.rice.edu/software/ARPACK/). We study glasses with $T_p$ ranging from well above the onset of
supercooling, denoted as  $T_p = \infty$, down to $T_p = 0.062$, which is about $60\%$ of the mode-coupling
temperature $T_c \approx 0.108$ ~\cite{swapPRX2017}. The onset of slow dynamics in an equilibrated fluid occurs around $T_o = 0.2$.
The parent temperature $T_p = 0.062$ is lower than the estimated experimental glass temperature $T_g \approx 0.072$
for this model~\cite{swapPRX2017}, and thus the glass with $T_p < 0.072$ qualifies as ultrastable.
The particle number $N$ varies between $48000$ and  $1000000$ for glasses  at $T_p=\infty$, and between $48000$ and $192000$ for glasses with
$0.062 \leq T_p \leq 0.120$.
For all glasses studied the number
density $\rho = 1.0$.

\subsection{Sound attenuation}

\begin{figure}
\centering
\includegraphics[width=0.5\textwidth]{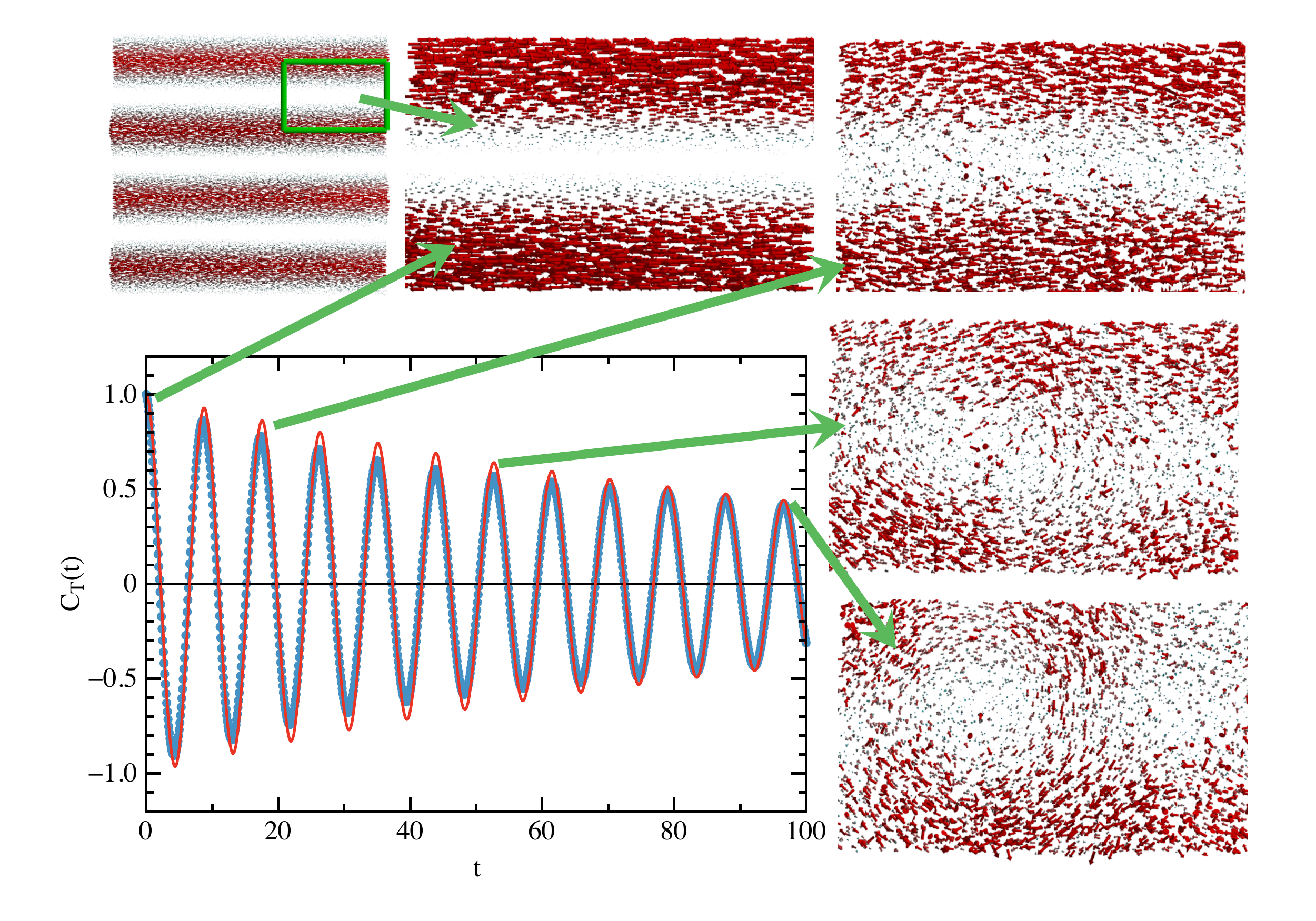}
\caption{  Decay of $C_T(t)$ for
a transverse excitation with a wavevector $\mathbf{k} = (0,4 \pi/L,0)$ for our most stable
glass, $T_p = 0.062$. The red curve is a fit to $C_T(t) = \exp(-\Gamma_\lambda t/2) \cos(\Omega_\lambda t)$.
The velocity field for the whole system is shown in the upper left corner and
for a section at representative times corresponding to the peaks in $C_T(t)$ indicated by the arrows.}
\label{fig1}
\end{figure}

\begin{figure*}
\centering
\includegraphics[width=0.35\textwidth]{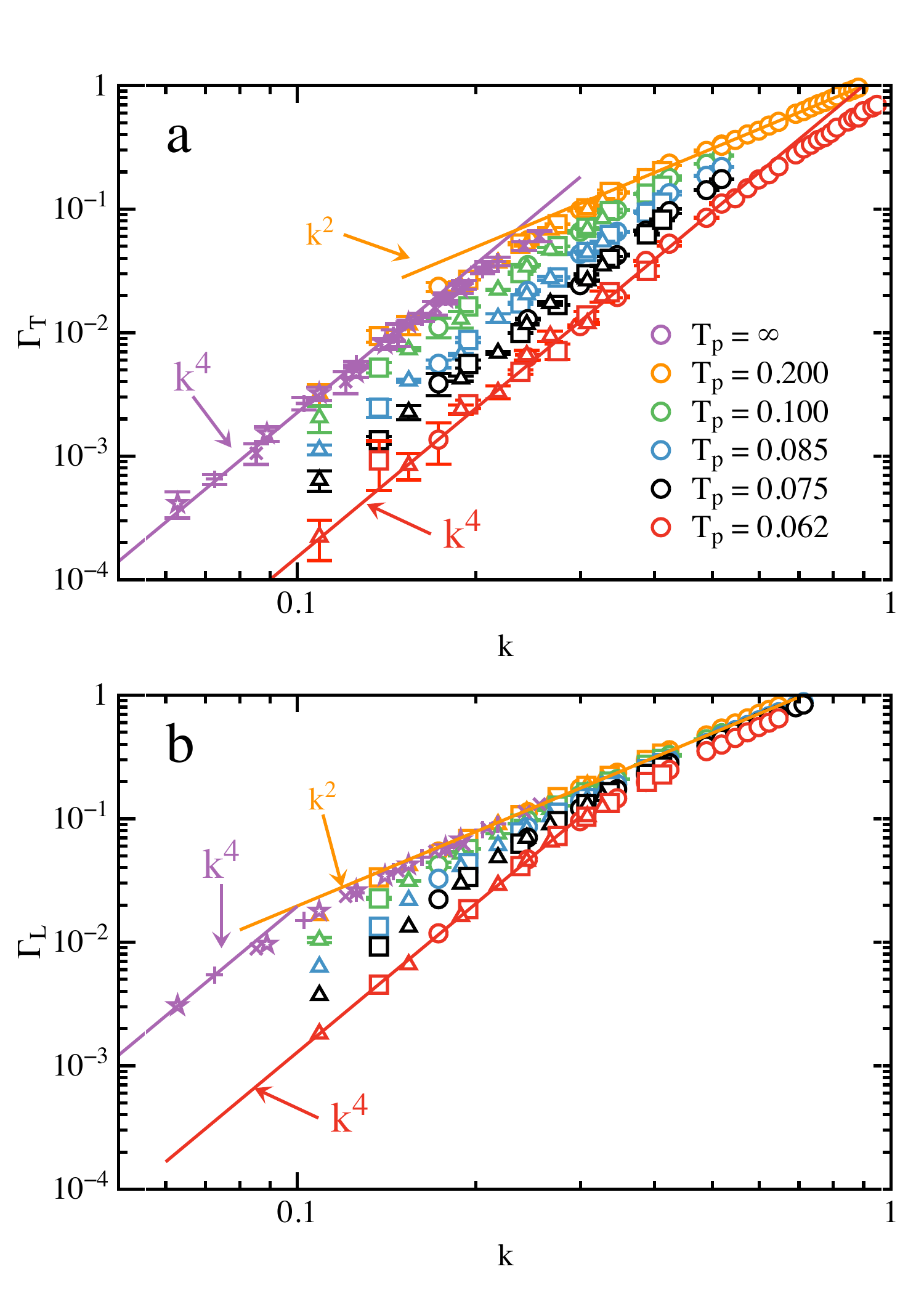}
\includegraphics[width=0.35\textwidth]{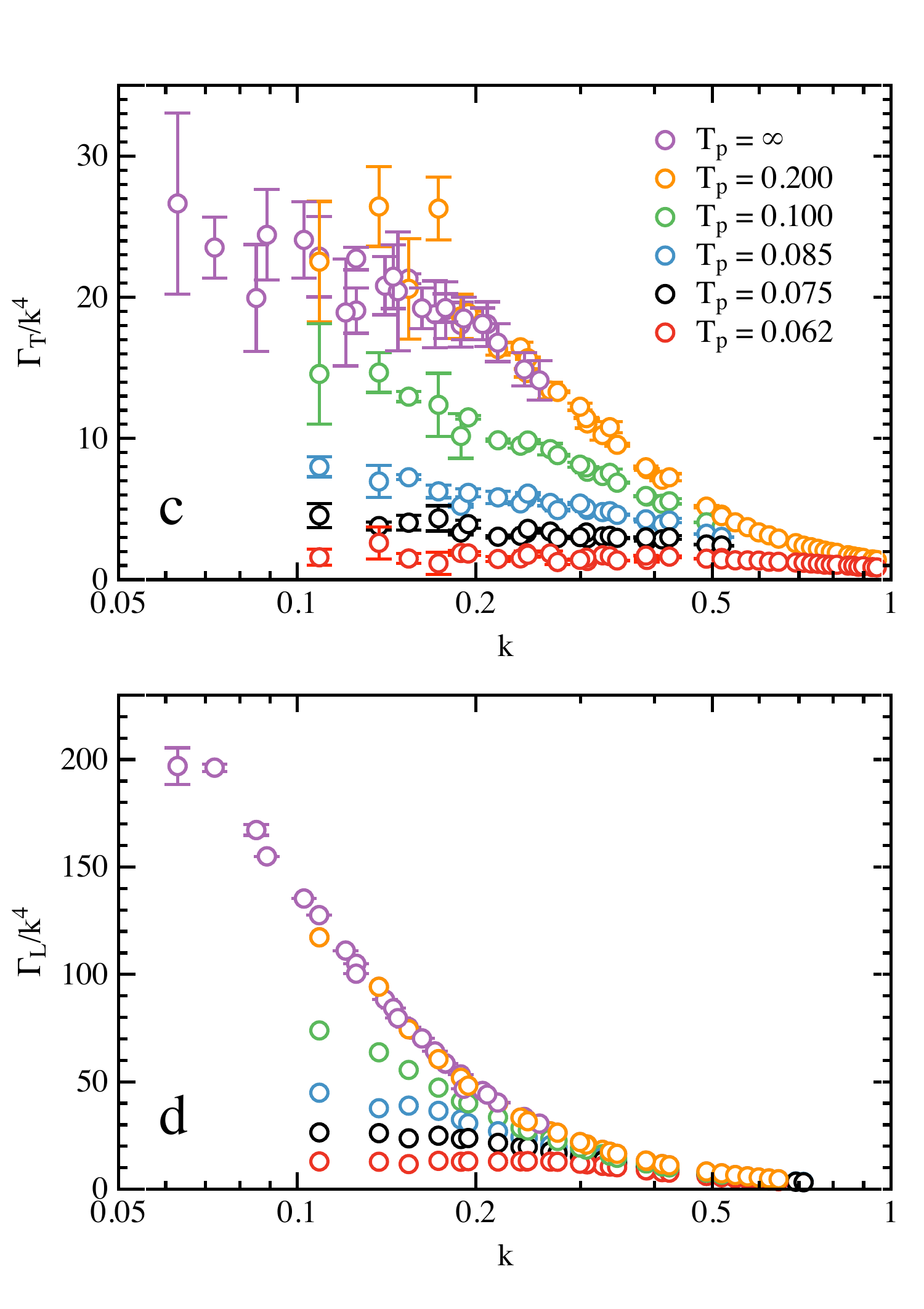}
\caption{ Wavevector $k$ dependence of
sound attenuation (a) $\Gamma_T(k)$ and (b) $\Gamma_L(k)$ in from poorly annealed
glasses ($T_p = \infty$) to stable glasses ($T_p = 0.062$).
The different symbols denote different system sizes: star=1000K, plus=600K, x=450K, triangle=192K, square=96K, circle=48K.
The $k^2$ dependence is evident at large wavevectors and the crossover to $k^4$ scaling can be seen for
$T_p = \infty$ and $T_p = 0.062$. The reduced sound attenuation (c) $\Gamma_T/k^4$
and (d) $\Gamma_L/k^4$. A straight line with negative slope would indicate a logarithmic correction,
which is valid only for a small range of wavevectors. }
\label{fig2}
\end{figure*}

We use two different methods to obtain sound attenuation:1) we calculate the $T=0$ dynamic structure factor utilizing the eigenvalues and eigenvectors of the
dynamic matrix \cite{method_only_WangPRL2015};   2) we study the decay
of an excited sound wave in the harmonic approximation \cite{GelinNatMat2016}.

We calculate the $T=0$ dynamic structure factors using the eigenvalues and eigenvectors of the
dynamic matrix \cite{method_only_WangPRL2015},
\begin{equation}
 S_\lambda(k,\omega) = \left(\frac{k^2}{N \omega^2}\right)
\sum_{n=1}^{3N-3} F_{n,\lambda}(k)\delta(\omega-\omega_n),
\label{sk}
\end{equation}
where $\lambda$ is $T$ for transverse or  $L$ for longitudinal structure factor,
$\omega_{n}$ is the frequency (square root of the eigenvalue) associated with the $n$-th eigenvector.
The sum is taken over all but the three modes corresponding to a universal translation.
In eqn~\ref{sk}
\begin{equation}
F_{n,T}(k) =
\left|\sum_{j=1}^{N}(\mathbf{e}_{n,j}\times \hat{\mathbf{k}})e^{i \mathbf{k}\cdot \mathbf{r}_{j}^0}\right|^2,
\label{eq4}
\end{equation}
and
\begin{equation}
 F_{n,L}(k) =
\left|\sum_{j=1}^{N}(\mathbf{e}_{n,j} \cdot \hat{\mathbf{k}})e^{i\mathbf{k}\cdot\mathbf{r}_{j}^0}\right|^2,
\label{eq5}
\end{equation}
where  $\mathbf{e}_{n,j}$ is the polarization vector of particle $j$ in the  $n$-th eigenvector,
$\mathbf{r}_{j}^0$ is the position of  particle $j$ in the inherent structure, $\mathbf{k}$ is the wavevector satisfying periodic
boundary conditions,  $k\equiv |\mathbf{k}|$  and $\hat{\mathbf{k}}=\mathbf{k}/|\mathbf{k}|$. We extract the damping
coefficients $\Gamma_{\lambda}$ and the characteristic frequencies $\Omega_{\lambda}$ by fitting $S_\lambda(k,\omega)$ to a damped harmonic oscillator model \cite{Rayleigh_MonacoPNAS2009_sim},
\begin{equation}
S_\lambda(k,\omega) \propto
\frac{\Omega^2_{\lambda}(k)\Gamma_{\lambda}(k)}
{[\omega^2-\Omega^2_{\lambda}(k)]^2+\omega^2\Gamma^2_{\lambda}(k)}.
\label{dampedharmonic}
\end{equation}


Another method to determine $\Gamma_\lambda$ and $\Omega_\lambda$ is to study the decay of excited sound waves in the harmonic approximation, and most of our  results shown in this work  are from this method (unless specified). Specifically, following Ref.~\cite{GelinNatMat2016}, we excite a sound wave at $t=0$ by giving each particle a velocity
$\dot{\mathbf{u}}_i^0 = \mathbf{a}_\lambda \sin(\mathbf{k} \cdot \mathbf{r}_i^0)$, where
$\mathbf{a}_L \propto \hat{\mathbf{k}}$
and $\mathbf{a}_T \cdot \mathbf{k} = 0$.  We then
numerically solve the equations of motion,
\begin{equation}
\ddot{\mathbf{u}}_{i}(t)=  -\sum_{j=1}^{N} \mathbf{D}_{ij} \cdot \mathbf{u}_{j}(t) + \dot{\mathbf{u}}_i^0 \delta(t)
\label{Harmonic}.
\end{equation}
Here, $D_{ij}$ is  dynamic matrix and $\mathbf{u}_i(t)$ denotes the displacement of particle $i$  at $t$ from its  inherent structure position.
We calculate the velocity correlation function,
   \begin{equation}
 C_{\lambda} (t) =
\frac{ \sum_{i=1}^{N} \dot{\mathbf{u}}_i(0) \cdot  \dot{\mathbf{u}}_i(t)}
{\sum_{i=1}^{N} \dot{\mathbf{u}}_i(0) \cdot  \dot{\mathbf{u}}_i(0) } ,
\label{velo}
\end{equation}
and fit it to
\begin{equation}
C_{\lambda} (t)=\exp{(-\Gamma_{\lambda}(k)t/2)}\cos(\Omega_{\lambda}(k)t),
\end{equation}
to determine the frequency $\Omega_\lambda$ and the sound attenuation $\Gamma_\lambda$.
Since the calculation obtains $\Omega_\lambda$ through a fit for a fixed $\mathbf{k}$, the wavevector
is precisely known but there is uncertainty in $\Omega_\lambda$.

Shown in Fig. \ref{fig1} is an example of the excited sound wave method~\cite{GelinNatMat2016}. The snapshots in Fig. \ref{fig1} show
the velocity field for $\mathbf{k} = (0,4 \pi/L,0)$ in a 48000 particle system for
times at the peak values of $C_T(t)$ indicated in the figure. As expected, the  sound wave is scattered
and the initial velocity profile decays.

The
two methods introduced above encode the same dynamical information, but there exists a finite size effect that is impossible to correct for using the
normal mode analysis. See \textbf{Appendix} section for details on how we account for this finite size effect and for details on how we obtain
$\Gamma_\lambda$.

\section{Sound attenuation in stable glasses}

Shown in Fig. \ref{fig2} are $\Gamma_\lambda(k)$ for a range of stabilities for
(a) transverse sound waves and (b) longitudinal sound waves.
For large wavevectors we observe quadratic scaling, which is consistent
with previous results. There is no difference in the
attenuation for $T_p = 0.2$ and $T_p = \infty$ suggesting
that zero-temperature glasses quenched from parent temperatures above the onset temperature $T_o = 0.2$ have
identical attenuation. There is a crossover to quartic scaling, Rayleigh scaling,
for our least stable glasses $T_p = \infty$ and our most stable glasses
$T_p = 0.062$.  Therefore, $\Gamma_\lambda(k) = B_\lambda k^4$
for small wavevectors irrespective of the glass's stability.

To examine the stability dependence of
$B_\lambda$ and the possibility of a logarithmic correction, in Fig. \ref{fig2} we plot
$\Gamma_\lambda(k)/k^4$ for
$T_p = \infty$, 0.1, 0.085, 0.075, and 0.062 for
transverse sound (c) and the longitudinal sound (d). There
is a factor of 15 decrease in $B_\lambda$ from our least stable glass to the most stable glass.
We note that in the representation of Fig. \ref{fig2} a straight line with a negative slope would indicate the
$-k^4 \ln(k)$ scaling suggested
by Gelin \emph{et al.}
~\cite{GelinNatMat2016}.  We can identify a range of wavevectors that is described by
$\Gamma_T(k) \sim - k^4 \ln(k)$ for our least stable glasses, but this does not provide a good
description for a wide range of wavevectors.
Instead we observe a distinct plateau at low wavevectors, indicating a purely quartic scaling without a logarithmic correction.

\begin{figure}
\centering
\includegraphics[width=0.40\textwidth]{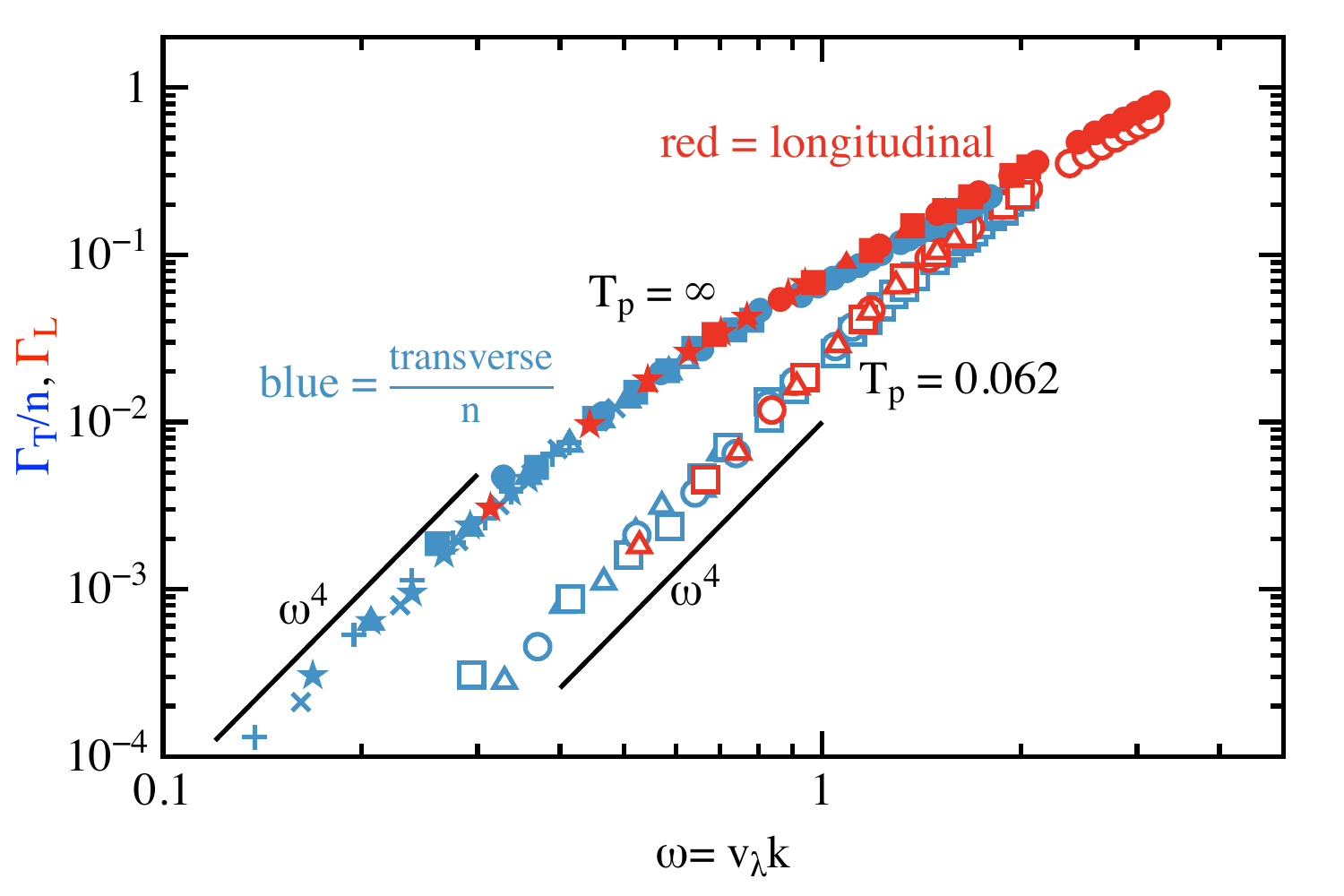}
\caption{ Frequency $\omega = v_\lambda k$ dependence of
sound attenuation for our least stable glass $T_p = \infty$ (filled symbols) and our most stable glass $T_p = 0.062$ (open symbols).
The different symbols denote different system sizes: star=1000K, plus=600K, x=450K, triangle=192K, square=96K, circle=48K.
The red symbols are the results for the longitudinal attenuation and the blue symbols are results for the transverse attenuation.
The transverse attenuation is scaled by a $T_p$ dependent factor $n$, where $n=5$ for $T_p = \infty$ and
$n = 3$ for $T_p = 0.062$.  }
\label{GS}
\end{figure}

As noted by Monaco and Mossa \cite{Rayleigh_MonacoPNAS2009_sim} when studying glasses created by quenching from  mildly supercooled liquids,   the transverse and longitudinal sound
attenuation differ by a constant factor when examined as a function of frequency $\omega = v_\lambda k$,
where $v_T = \sqrt{G/\rho}$, $v_L = \sqrt{(K + 4G/3)/\rho}$, $G$ is
the shear modulus, and $K$ is the bulk modulus, Fig.~\ref{GS}. We find $\Gamma_L(\omega)= \Gamma_T(\omega)/n$ irrespective of the glass's stability, but the scaling factor $n$ is stability dependent with $n \approx 5$ for our poorly annealed glass, $T_p = \infty$,
and $n \approx 3$ for our most stable glass, $T_p = 0.062$, indicating a decreasing difference between $\Gamma_T(\omega)$ and $\Gamma_L(\omega)$  with increasing stability.   This scaling suggests that the sound
attenuation is governed by a stability dependent frequency (time) scale and possibly not a
characteristic length scale. However, a changing length scale cannot be ruled out.

With increasing stability, the glass becomes less dissipative and quartic scalings of $\Gamma_T$ and $\Gamma_L$ start at larger
wavevectors. The wavevector at which the quartic scaling begins depends on the polarization, transverse or longitudinal, of the sound wave.
In contrast, if we plot the sound attenunation as a function of frequency, the frequency where the quartic scaling begins
does not depend on the transverse or longitudinal sound wave. Again, this crossover frequency increases with increasing stability.
The glass is becoming more uniform, resulting in a decrease in the dissipation
~\cite{GelinNatMat2016,John1983} with an increase in the stability.

\begin{figure}
\centering
\includegraphics[width=0.40\textwidth]{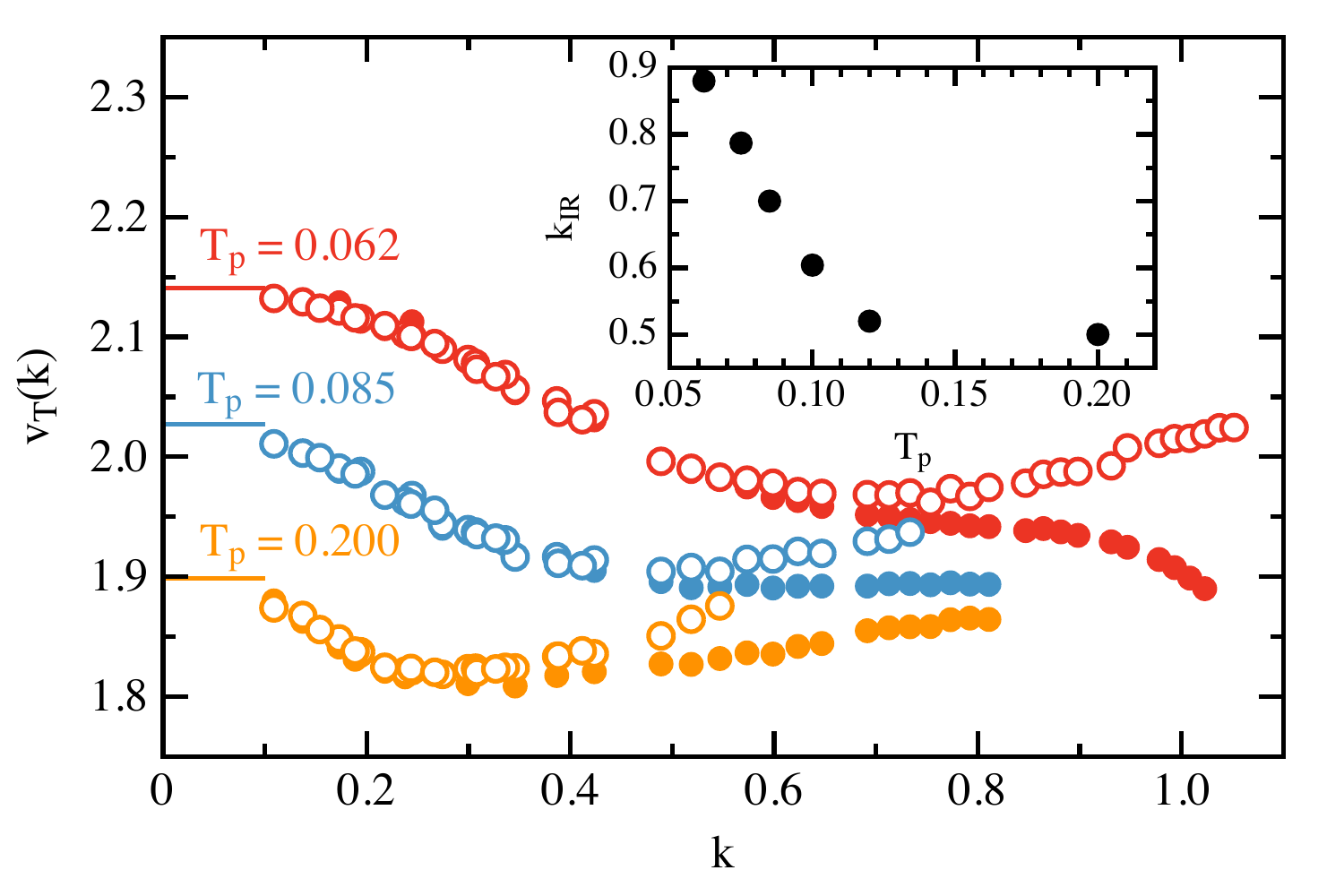}
\caption{
 The wavevector dependence of sound speed for different parent temperatures $T_p$. The  horizontal  lines indicate the corresponding macroscopic
values in the long-wavelength limit. The open symbols are obtained through fits of $C_T(t)$ and the closed symbols are
obtained through fits to $S_T(k;\omega)$.  (Inset)  Ioffe-Regel wavevector $k_{IR}$ as a function of  $T_p$.   }
\label{fig4}
\end{figure}

For small and intermediate wavevectors the wavevector-dependent speed of sound
$v_T(k) = \Omega_T/k$ 
is a well
defined quantity. In particular, for every parent temperature the $k \rightarrow 0$ limit is
given by $\sqrt{G/\rho}$, which is shown as horizontal lines in Fig. \ref{fig4}.
However, with increasing wavevector different methods lead to slightly but systematically different results for the
wavevector-dependent speed of sound. If we determine the speed of sound from the fit to the frequency-dependent dynamic structure
factor (filled circles), the resulting quantity exhibits a minimum, which has been reported in
previous simulations \cite{MizunoArxiv2018phonontransport,Rayleigh_MonacoPNAS2009_sim,GelinNatMat2016,MarruzzoScientific2013}
and experiments \cite{Rayleigh_BaldiPRL2010,Rayleigh_MonacoPNAS2009_exp,WangPRB2018}. This minimum is replaced
by a plateau for our stable glasses. However, if we rely upon the fit to the time-dependent function $C_\lambda(t)$ (open symbols),
the wavevector-dependent speed of sound exhibits a more pronounced minimum, which is also present for the stable glasses.
The difference between the two methods is small but systematic.

It would be expected that the two methods could disagree when the excitation is no longer well described as a propagating
sound wave, which is generally associated to when the mean free path is equal to half the wavelength, \emph{i.e.}\ the Ioffe-Regel limit.
Shown in the inset to Fig.~\ref{fig4} is the Ioffe-Regel limit obtained from when $\Omega_T(k_{IR}) = \pi \Gamma_T(k_{IR})$
as a function of the parent temperature.
For this calculation we used $\Omega_T$ determined from the fits to the dynamic structure factor. The result is not sensitive
to which method is used to determine $\Omega_T$.
For $T_p = 0.2$, $k_{IR} \approx 0.5$ and for $T_p = 0.062$, $k_{IR} \approx 0.87$. Both of these quantities lie slightly above
where the two methods to obtain the wavevector-dependent speed of sound begin to diverge. Thus, the classification
of these excitations as propagating sound waves is breaking down for wavevectors slightly smaller than $k_{IR}$.

Nevertheless, we find that increasing the stability of the glass allows for
propagating sound waves at smaller wave lengths, and this can be quantified
by the change in $k_{IR}$. For decreasing $T_p$, $k_{IR}$ increases by a factor of 1.8 over our range of stability.
For wavevectors above $k_{IR}$
it is expected that the vibrations are more localized and there is a change in the energy transport from a propagating regime
below $k_{IR}$ to a diffusive regime above $k_{IR}$ \cite{Beltukov2018,Xu2009,Vitelli2010PRE}.  Therefore, the decreased dissipation
and increase in $k_{IR}$ should have significant effects on the thermal conductivity and the stability
dependence of thermal energy transport.

\section{Connection between sound attenuation, vibrational modes, and the boson peak}

A recurring idea is that sound attenuation and the excess in vibrational modes over the
Debye theory are intimately connected.  Recall that in the Debye theory the density of states increases
with a decrease in the speed of sound.
Using this idea, the minimum in $v_T(\omega)$ has been associated with an increase of the density of states $D(\omega)$ and the
boson peak  using a generalized plane wave approach \cite{Rayleigh_MonacoPNAS2009_sim}. However, we find that the description of the vibrational modes as well defined sound waves breaks down for wavevectors below the
boson peak whose position approximates $k_{IR}$ \cite{Tanaka2008NM}.


A generalized Debye model of Mizuno and Ikeda \cite{MizunoArxiv2018phonontransport} and the theoretical treatment
of Schirmacher \textit{et al.} \cite{Schirmacher2007} both relate the excess number of low frequency modes above the Debye model,
$D_{\mathrm{ex}}(\omega)$, to sound attenuation. Both of these treatments
predict that $D_{\mathrm{ex}} \approx 4 B_T/(\pi k_D^2 v_T^6) \omega^4$,
where $k_D = (6 \pi \rho)^{1/3}$, for small wavevectors. In previous studies \cite{Mizuno2017,WangarXiv2018} it was
found that the low frequency modes could be divided into extended and
localized modes. The density of the low frequency extended modes obeys Debye theory and
the density of the localized modes $D_{\mathrm{loc}} = A_4 \omega^4$.
Therefore, these localized modes are the modes in excess of the Debye theory, and
we can associate them with $D_{\mathrm{ex}}(\omega)$.
We find that $A_4$ is  20\% larger than $4 B_T/(\pi k_D^2 v_T^6)$
for our poorly annealed glass and 150\% larger for our most stable glass.
Therefore, these models are currently not quantitatively predictive and get worse with increasing
stability.

Our findings for the transverse sound attenuation in moderate and low stability glasses are in general agreement with
the very recent results of Moriel \textit{et al.}~\cite{Lerner2019}. Specifically, both our study and
that of Moriel \textit{et al.}\ find quartic small wavevector scaling of the transverse sound attenuation in 3D glasses.
Moriel \textit{et al.}\ also investigated the dependence of the sound attenuation
of glasses with different densities
of low-frequency quasi-localized modes. 
They found that the decreasing density of these modes
correlates with the decreasing extent of the intermediate regime between the small
wavevector quartic scaling and the large wavevector quadratic scaling, which can be fitted to the $-k^4\ln k$ form
proposed by Gelin \textit{et al.} \cite{GelinNatMat2016}.

\begin{figure}
\centering
\includegraphics[width=0.40\textwidth]{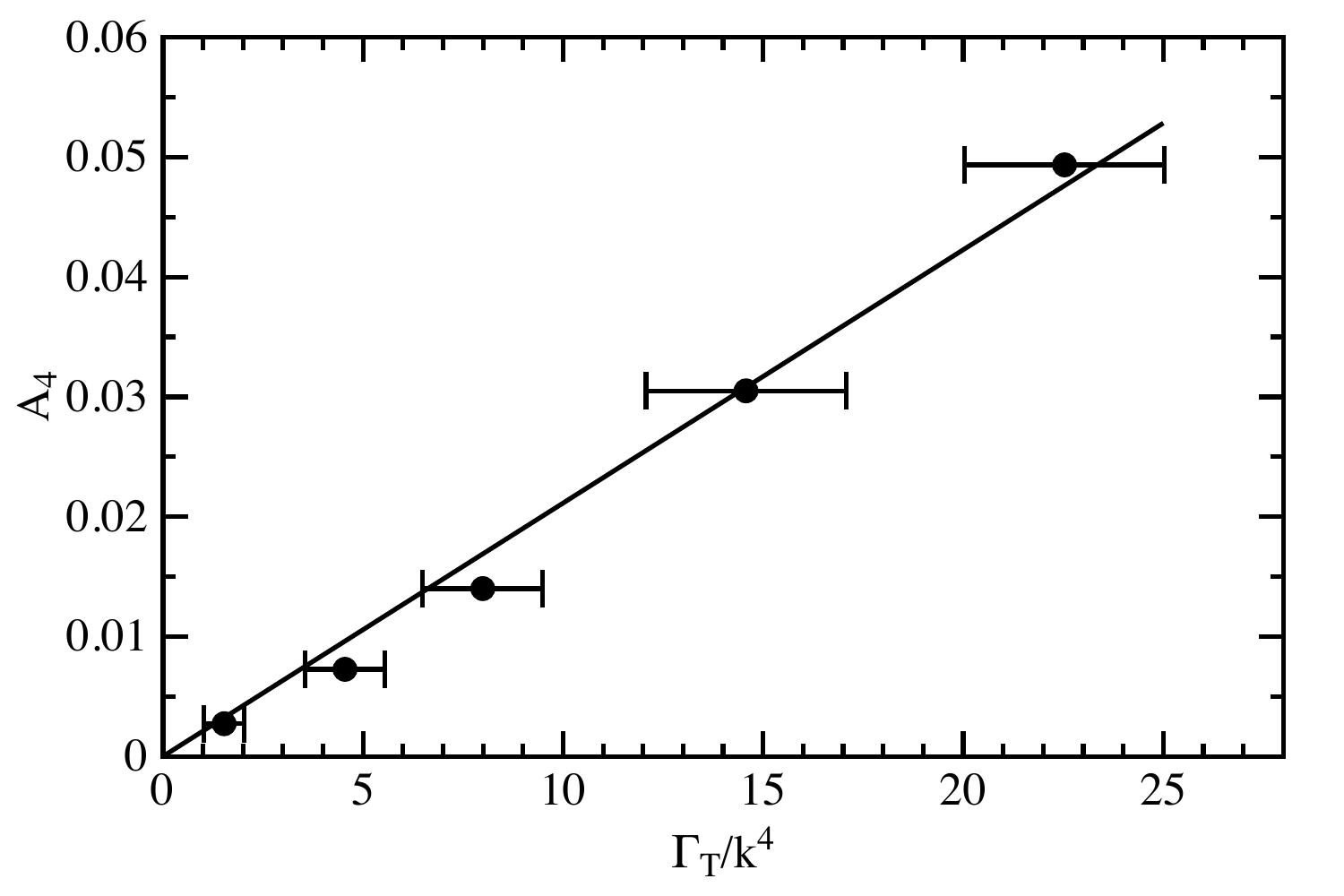}
\caption{ The
coefficient $A_4$ describing the the density of low frequency quasi-localized modes,
$D_{\mathrm{loc}} = A_4 \omega^4$ correlates very well with the plateau height of $\Gamma_T/k^4$ for small wavevectors. They are both strongly suppressed when glass stability increases.}
\label{plateau}
\end{figure}

In contrast, we investigated a number of glasses with very different stabilities. 
We find that the sound attenuation and the Rayleigh scattering plateau $\Gamma_\lambda/k^4$  decrease rapidly.
In Fig.~\ref{plateau} we show that $A_4$ and $B_T=\Gamma_T/k^4$ are proportional to each other,
with $B_T \sim A_4$.
This significantly extends the qualitative correlation found
by Moriel \textit{et al.}~\cite{Lerner2019}

A recent experiment by Pogna \textit{et al.}\ \cite{Pogna2019} reported on a connection between sound
attenuation and the boson peak. They find a decrease in the
boson peak height and sound attenuation for hyperaged amber (conjectured to be much more stable) compared to
annealed amber (with ordinary stability),
which mirrors our results \cite{WangarXiv2018}.  Pogna \textit{et al.}\ used
the fluctuating elasticity theory of Schirmacher \textit{et al.} \cite{Schirmacher2010}
(which predicts the quartic scaling of sound attenuation with the wavevector) to fit the vibrational density of states.
There are two main parameters in
the theory, one quantifies the strength of the disorder and is
related to the width of the local elastic constant distribution, and
another quantifying the spatial range of correlations of elasticity.  They concluded that upon lowering the fictive
temperature by 9\% that there was a six percent decrease of the strength of the fluctuations
and a 22\% increase of the elastic correlation length.
Therefore, they conjectured that the change of the low-frequency vibrational properties is mainly driven by an
increased elastic correlation length.
Future work should examine the change of the disorder
strength and the elastic correlation length with stability more directly to verify this conclusion.

A competing theoretical explanation for the relationship between sound attenuation and the boson peak
is that the sound modes interact with additional soft modes \cite{Buchenau1992}, the
soft potential model. Examination and evaluation of the soft potential model requires the determination
of several parameters, and this exercise is left for future work.

\section{Discussion}

The idea that a Rayleigh scattering mechanism may be responsible for the small wavevector
scaling of sound attenuation spans for over 60 years \cite{Klemens1951,MizunoArxiv2018phonontransport}.
Mizuno and Ikeda considered scattering of an elastic wave. Their analysis determined that
$\Gamma_\lambda = \delta \gamma_\lambda^2 D_{\lambda}^3 \Omega_\lambda^4/(4 \pi v_\lambda^3)$, where
$\delta \gamma$ is the strength of the elastic inhomogeneities and $D$ is the characteristic
size \cite{MizunoArxiv2018phonontransport}. Since it has been suggested that $k_{BP} = \omega_{BP}/v_T$ is related to the
length scale of elastic inhomogeneities \cite{Schirmacher2015,WangPRB2018}, and thus $D$, we checked to see if this was consistent
with the quartic scaling regimes for $\Gamma_T$. We used the thermodynamic approach
studied by Mizuno, Mossa, and Barrat \cite{Mizuno2013} to obtain the strength $\delta \gamma_T = \delta G/G$,
where  $\delta G$ is the fluctuations of the shear modulus, of the elastic inhomogeneities.
We find that this naive approach does not correctly predict the change in the sound attenuation for each
parent temperature. Future work needs to examine the spatial correlations of the elastic modulus and their
relationship to sound attenuation.

Recent experiments on amber aged for 110 million years suggest that the vibrational properties of amorphous materials are controlled
by the distribution of elastic constants and their spatial correlation \cite{Pogna2019}. Future numerical studies should examine
this relationship for simulated ordinary and stable glasses to clarify this relationship.
The stability dependence of sound attenuation using
ultrastable glasses, experimentally available via the method of physical vapor deposition~\cite{SwallenScience2007},
has shown that sound damping decreases with increasing stability \cite{Pogna2015}. It would be interesting to examine the
wavevector dependence of sound damping at low temperatures where anharmonicities may come into play \cite{Mizuno2019damping}, for these ultrastable glasses.

\section*{Acknowledgements}
We thank E. Lerner and E. Bouchbinder for correspondence on an earlier version of this work. L.W., E. F., and G.S. acknowledge funding from NSF DMR-1608086. This work was also supported by a grant from the Simons foundation (No. 454933 L. B.). L. W., and P. G. acknowledge support from the National Natural Science Foundation of China (No. 51571011), the MOST 973 Program (No. 2015CB856800), and the NSAF joint program (No. U1530401). We acknowledge the computational support from Beijing Computational Science Research Center.

\section*{Appendix}
\begin{figure}
\centering
\includegraphics[width=0.40\textwidth]{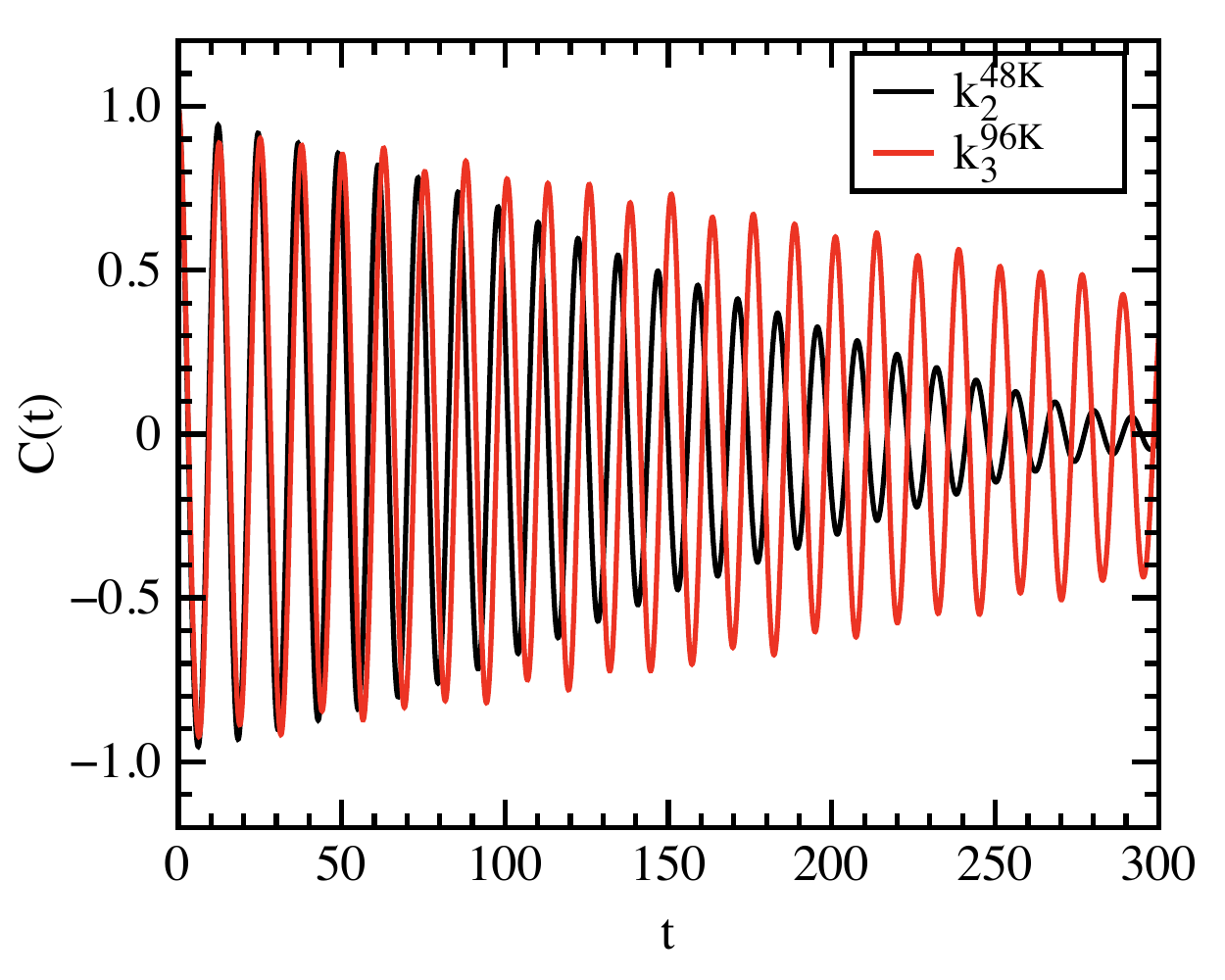}
\caption{\label{ctcompare}Velocity correlation function $C(t)$ for wavevectors of similar
magnitude, $k \approx 0.24$, for two different system sizes. The decay rate
is clearly different and does not appear exponential.}
\end{figure}

Molecular dynamics simulations can be subject to effects due to small size of the simulation cell
compared to experimental systems and the use of periodic boundary conditions. Bouchbinder and
Lerner recently commented on finite size effects in the calculation of the frequency width
of phonon bands \cite{Bouchbinder2018}, which indicates that finite size effects exist for
the calculation of sound attenuation in amorphous solids. We find that there are strong finite
size effects for the lowest wavevector sound waves in our simulations, especially
for our most stable glasses. Here we describe
a method to calculate sound attenuation that is independent of system size.

One route to calculate the attenuation of sound waves is to study the decay of an
excitation in the harmonic approximation as described in the \textbf{Methods} section. After exciting a sound wave,
we study the decay of the velocity correlation function $C(t)$, eqn (\ref{velo}).
For small wavevectors we expect that $C(t) = \exp(-\Gamma_\lambda t/2)\cos(\Omega_\lambda t)$.

To demonstrate that a finite size effect exists we can examine $C(t)$ for
similar wavevectors in two systems of different sizes. The
magnitude of the third smallest allowed
wavevector for the 96K system $k_3^{96K} = 0.238$ and the magnitude of the
second smallest allowed wavevector is $k_2^{48K} = 0.245$. The attenuation
of these sound waves should be similar, but we find that they are very different, Fig.~\ref{ctcompare}.
Specifically, at long times the peak heights of the 96K system are much larger than for the 48K system.
However, $C(t)$ nearly overlaps at short times for both system sizes. To study the
decay of $C(t)$ we calculate the envelope of $C(t)$, which is the absolute value of the
maximum and minimum of the oscillations.

\begin{figure}
\centering
\includegraphics[width=0.40\textwidth]{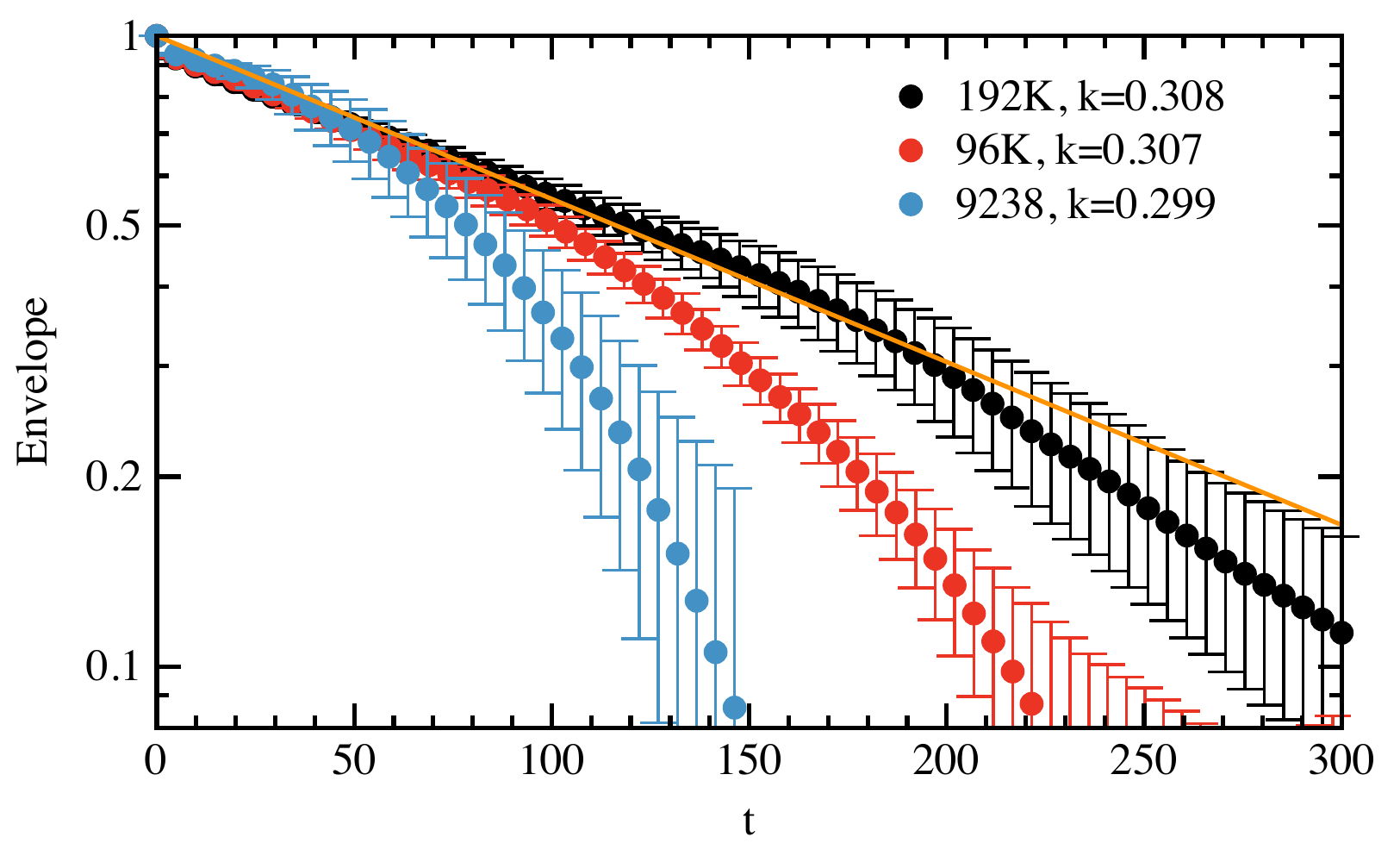}
\caption{\label{envelope} The envelope of of $C(t)$ for three system sizes
for three wavevectors of nearly equal magnitude. The solid line represents a fit of the envelope to $\exp(-\Gamma_\lambda t/2)$. }
\end{figure}

Shown in Fig.~\ref{envelope} on a linear-log scale is the envelope for three different sizes for a
wavevector of similar magnitude. We note that the initial decay of all three envelopes is exponential,
but there are deviations from the exponential decay at a system size dependent time. To
determine $\Gamma_\lambda$ we fit the envelope to $\exp(-\Gamma_\lambda t/2)$ up to a time
when the decay is no longer exponential. Our uncertainty in $\Gamma_\lambda$ reflects the
uncertainty in this fitting range.

Another method to obtain sound attenuation is through the dynamic structure factor
$S_\lambda(k,\omega)$ using the eigenvalues and eigenvectors of the dynamic  matrix,
as described in the \textbf{Methods} section, or Fourier transforming $C(t)$. Sound
attenuation $\Gamma_\lambda$ is then obtained by fitting with the damped harmonic oscillator model, eqn (\ref{dampedharmonic}).

\begin{figure}
\centering
\includegraphics[width=0.40\textwidth]{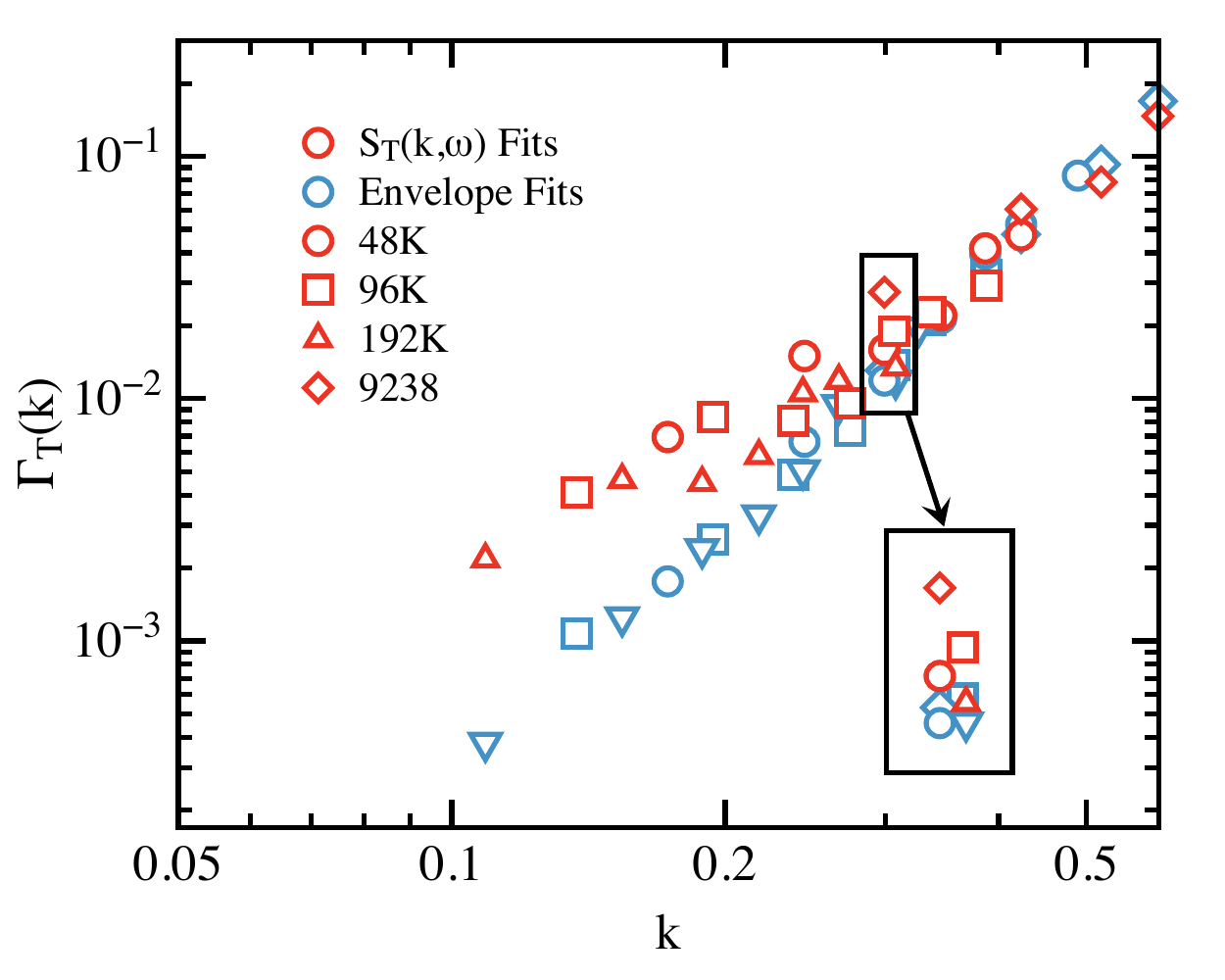}
\caption{\label{compare}Sound attenuation $\Gamma_T(k)$ calculated using fits to $S_T(k,\omega)$ (red) and
the envelope fits (blue). The different symbols correspond to different system sizes. The inset shows an expanded
view of the results for one wavevector. There is a clear finite size effect when $\Gamma_T(k)$ is obtained by
fitting $S_T(k,\omega)$, which can be removed by using the restricted envelope fits.}
\end{figure}

Shown in Fig.~\ref{compare} as red symbols are the results of fitting $S_T(k,\omega)$
and as blue symbols are the results of the restricted envelope fits. The different symbols indicate
different system sizes.  The inset shows an expanded view of a region of very similar wavevectors for
four different system sizes.
There is a clear finite size effect when $\Gamma_T$ is found through fits of $S_T(k,\omega)$, which
is removed by using the envelope fits.

\end{document}